\documentclass[aps,amsmath,amssymb,preprint,superscriptaddress]{revtex4}
\usepackage{graphicx}
\usepackage{dcolumn}
\usepackage{bm}
\usepackage{latexsym,bm,amsmath,amssymb}
\usepackage{textcomp}
\usepackage{gensymb}

\begin{document}

\title{Transformation Hydrodynamic Metamaterials: Rigorous Arguments on Form Invariance and Structural Design with Spatial Variance}

\author{Gaole Dai}\email{gldai@ntu.edu.cn}
\affiliation{School of Sciences, Nantong University, Nantong 226019, China}

\author{Jun Wang}\email{wj21@ecust.edu.cn}
\affiliation{School of Physics, East China University of Science and Technology, Shanghai 200237, China}
\affiliation{Wenzhou Institute, University of Chinese Academy of Sciences, Wenzhou 325001, China}

\begin{abstract}
The method of transformation optics has been a powerful tool to manipulate physical fields if governing equations are formally invariant under coordinate transformations. However, regulation of hydrodynamics is still far from satisfactory due to the lack of rigorous arguments on the validation of transformation theory for various categories of fluids. In this paper, we systematically investigate the applicability of transformation optics to fluid mechanics. We find that the Stokes equation and the Navier-Stokes equations,  respectively describing the Stokes flow and general flow, will alter their forms under curvilinear transformations. On the contrary, the Hele-Shaw flow characterized with shallow geometries rigidly retain the form of its governing equation under arbitrary transformations. Based on the derived transformation rules, we propose the design of multilayered structures with spatially varying cell depth, instead of engineering the rank-2 shear viscosity tensor, to realize the required anisotropy of transformation Hele-Shaw hydrodynamic metamaterials. The theoretical certify and fabrication method revealed in this work may pave an avenue for precisely controlling flow distribution with the concept of artificial structure design.

\end{abstract}

\maketitle

\section{Introduction}

Transformation optics (TO), based on the form invariance of Maxwell's equations under coordinate transformations, has triggered the discovery of many exotic phenomena and rapid progress of metamaterials in electromagnetism~\cite{Science06-01,Science06-02,siam,NM2010,laser}. This geometric approach to mimic physics in curved space has also been successfully applied to other wave or diffusion systems~\cite{rpp,AM2021}. It is notable that not all physical phenomena satisfy the form invariance required by TO. For example, Milton \textit{et al.} found that the form of the conventional elastodynamic equation  with a rank-4 elasticity tensor actually changes  unless certain modifications are added~\cite{NJP2006}.  However, such a rigorous treatment on the validation of TO is still absent in another fundamental topic of continuum theories, the fluid mechanics,  
whose  governing equations include  the  Navier-Stokes (NS) equations known for its complexity~\cite{landau-6}. 
Early works on transforming fluid mechanics took different ways to simplify the NS equations  and   realized functions like invisibility in underwater acoustic waves~\cite{NJP2007,APL2007,PRB2017}, liquid surface waves~\cite{epl09,prl2022hy,prl08} and Darcy flows in porous media~\cite{PRL2011,PRE2018}. Recently, a series of transformation hydrodynamic metamaterials~\cite{PRL2019,EML2020-2,PRAP2019}  were experimentally demonstrated in Stokes or creeping flows with  anisotropic  inhomogeneous  viscosities, built upon the form invariance of the Stokes  equation  (neglecting the inertial term in the NS equations). Later, this approach  was extended to general laminar flows~\cite{EML2020-1},  which assumes the NS equations are also form-invariant. Although the recent works appear to greatly expand the applicability of TO for fluid manipulation, we find that there are still some problems with the strictness of the theory that cannot be ignored.

First, the general viscosity  should be a rank-4 tensor, while the recent works~\cite{PRL2019,EML2020-2,PRAP2019,EML2020-1} rely on the assumption of a rank-2 tensor as an anisotropic version of scalar shear viscosity.  We have not yet checked whether it can be reconciled to realistic viscous constitutive relation. Second, even if the viscosity can be reduced to a rank-2 tensor, the Stokes flows still might not meet the requirements of TO, let alone general (laminar) flows~\cite{pd}. In fact, we prove that  neither  the Stokes equation nor the NS equations  can satisfy the form invariance in this work with both theoretical and numerical demonstrations.
However, recent works~\cite{PRL2019,EML2020-2,PRAP2019,EML2020-1} do give valid evidences that hydrodynamic metamaterials with an effective rank-2 viscosity tensor could work under certain conditions, both numerically and experimentally.
Therefore, we also aim to  find the conditions under which the TO theory can hold in fluid mechanics. In fact, some studies~\cite{arxiv19,PR2021,arxiv20,PRL2021} have noticed that Refs.~\cite{PRL2019,PRAP2019,EML2020-2,EML2020-1} actually used shallow geometries, i.e., the Hele-Shaw cells~\cite{nature1898}, which restrict the flows inside two plates (usually parallel) with a small gap. Hele-Shaw cells or Hele-Shaw flows are widely used in many fields like microfluidics~\cite{ARFM2004,AC2002}. We further prove that the Stokes equation in Hele-Shaw cells can be reduced to a form that satisfies TO when considering anisotropy and inhomogenity.
 In addition, to realize hydrodynamic metamaterials in Hele-Shaw cells, we propose an approach of transforming the depth of the cells instead of the viscosities. We achieve anisotropic depths required for transformation devices with engineered cell configurations in which layers with different depths are alternately arranged.

\section{Necessity of a shallow geometry for TO}

\subsection{Theoretical modeling}

In TO only involving spatial coordinate transformations, we consider a  map from the virtual space to the physical space. Usually, the virtual space is isotropic and homogeneous. TO requires the equation's forms in the two spaces are the same or the
form in the physical space can be an appropriate generalization for anisotropic inhomogeneous fluids.
In this work, we consider steady incompressible isothermal Newtonian fluid. The governing equations in the virtual space include the law of continuity
\begin{equation}\label{contunuity}
\nabla\cdot\left(\rho\mathbf v(\mathbf r)\right)=0,
\end{equation}
and the NS equations
\begin{equation}\label{NS}
\rho (\mathbf{v} \cdot \nabla)\mathbf{v}=-\nabla p+\nabla \cdot{{ \boldsymbol {\tau}}},
\end{equation}
where
$\rho$ accounts for the density, $\mathbf{v}$  is the velocity, $\mathbf r$  is the position vector,
$p$ is the pressure, and ${{ \boldsymbol {\tau}}}$ denotes
 the viscous stress tensor which can be expressed as
\begin{equation}\label{wbsl}
	{{ \boldsymbol {\tau}}}=\mu \left(\nabla \mathbf{v} +\nabla \mathbf{v} ^{\top }\right).
\end{equation}
Here $\mu$ is the  shear viscosity. For  Stokes flows, the inertial term $\rho (\mathbf{v} \cdot \nabla)\mathbf{v}$ can be neglected, and Eq.~(\ref{NS}) can be reduced to
\begin{equation}\label{Stokes}
	-\nabla p+\nabla \cdot {{ \boldsymbol {\tau}}}=0.
\end{equation}
It is notable that the viscosity is a scalar in Eq.~(\ref{wbsl}), which corresponds to isotopic incompressible fluids with some common symmetries~\cite{landau-6}.    It is easy to check that Eq.~(\ref{contunuity}) is form-invariant, and its  form in the physical space is
\begin{equation}\label{contunuity'}
	\nabla'\cdot\left(\rho'\mathbf v'(\mathbf r')\right)=0,
\end{equation}
where all the symbols are superscripted to represent quantities in the physical space. Let $\text{J}$ denote the Jacobian of the  map $\mathbf r \mapsto \mathbf r'$. The transformation rule to make Eq.~(\ref{contunuity'}) valid is~\cite{PR2021}
\begin{equation}\label{transv}
\mathbf v'(\mathbf r')=\text{J}(\mathbf r')\mathbf v(\mathbf r(\mathbf r'))\operatorname{det} \text{J}^{-1}(\mathbf r')
\end{equation}
since we can take density as a constant in incompressible isothermal flows.

Now, the question is whether Eqs.~(\ref{NS}) and (\ref{Stokes}) are also form-invariant as declared in Refs.~\cite{PRL2019,PRAP2019,EML2020-2,EML2020-1}. Their conclusions are based on a rank-2 viscosity tensor
\begin{equation}\label{vist}
\boldsymbol {\mu}'=\text{J}^{-\top}  \text{J}^{-1}\mu \det\text{J} 
\end{equation}
in the physical space, and the corresponding stress tensor is~\cite{PRL2019,PRAP2019,EML2020-2,EML2020-1} 
\begin{equation}\label{yl}
\nabla'\cdot\left({\boldsymbol {\mu}'}\cdot \left(\nabla' \mathbf{v}' +(\nabla' \mathbf{v}')^{\top }\right)\right).
\end{equation}
However, the general form of the viscosity is a rank-4 tensor denoted by $\mathbb{A}$ or $A_{ijkl}$ and the stress tensor should also be more complicated than Eq.~(\ref{yl})~\cite{landau-6}. When general anisotropy is introduced, $\mathbb{A}$ cannot be reduced to a rank-2 tensor; see our discussion in  Part A of Appendix. 
Even if such a rank-2 viscosity tensor does exist phenomenologically, whether Eqs.~(\ref{wbsl}) and (\ref{yl}) can  be transformed into each other should also be reconsidered.  
Anyway, the experimental results in  Refs.\cite{PRL2019,PRAP2019,EML2020-2} do give evidences that TO can work in some special flows, so the difference  between the observed systems and general Stokes flows or laminar flows should not be overlooked, i.e., the shallow geometries in experiments.

\begin{figure}[!ht]
	\centering
	\includegraphics[width=0.8\linewidth]{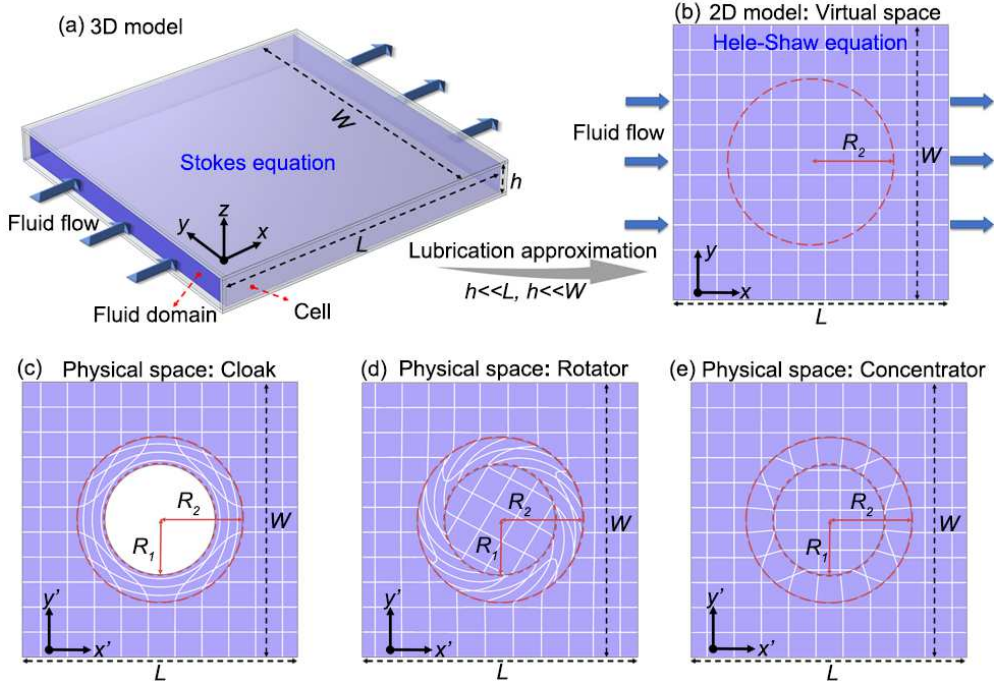}
	\caption{(a) Geometry of a 3D cell with a depth denoted as $h$. The top view of the cell is a rectangle with a length $L$ and a width $W$. The two $y$-$z$ boundaries (with regard to $x=\pm L/2$) are the flow inlet and outlet, respectively. Other four boundaries (with regard to $y=\pm L/2$ and $z=\pm h/2$) are nonslip walls.  (b) Geometry of a 2D Hele-Shaw cell in the $x$-$y$ plane. It also represents the virtual space for coordinate transformations to design  hydrodynamic metamaterials. The region outside the red dashed circle (with a radius $R_2$) undergoes the identity transformation. The region inside the circle can be transformed into  devices with different manipulation functions including: (c) cloak; (d) rotator; (e) concentrator. The transformation effects are also shown by the curved meshes.}\label{f1}
\end{figure}

To see the effect of shallow geometries, we consider flows in a cell or a cuboid-shaped pipe [Fig.~\ref{f1}(a)]. 
Its length and  width are $L$ and $W$, respectively. For simplicity, we take $L=W$ in the following discussion. $h$ denotes its height or the fluid depth if the cell is fully occupied. The inlet and outlet are on different open borders parallel to the $y$-$z$ plane.
The other four sides are solid walls with nonslip conditions.
Below we focus on the Stokes flows in  the cell. If the Stokes equation changes its form under coordinate transformations, so do the NS equations unless the inertial term can cancel out the form change of the stress tensor.  A shallow geometry means $h\ll L$ (e.g., $h/L=5\times 10^{-3}$ in Ref.~\cite{PRL2019}) and then the lubrication approximation can be used to obtain the velocity distribution for isotropic fluids:
$
\mathbf{v}=-\frac{h^{2}}{8 \mu}\left(1-\left(\frac{2 z}{h}\right)^{2}\right) \nabla p
$~\cite{InFlow}.
The average velocity in the $x$-$y$ plane (integrating along the $z$ axis with a symmetry to the $z=0$ plane) is~\cite{InFlow}
\begin{equation}\label{Hele}
\overline{\mathbf{v}}=\frac{2}{h} \int_{0}^{\frac{h}{2}} \mathbf{v} d z=-\frac{h^{2}}{12 \mu} \nabla p.
\end{equation}
If we take the average velocity as an approximation of $\mathbf{v}$, the three-dimensional (3D) model can be viewed as a two-dimensional (2D) one, i.e., the Hele-Shaw flow. Considering the law of continuity  [Eq.(\ref{contunuity})]  at the same time,
 the   governing equation of Hele-Shaw flows (Hele-Shaw equation) can be written as~\cite{InFlow} 
\begin{equation}\label{Helea}
\nabla\cdot\left(\frac{h^{2}}{12 \mu} \nabla p(\mathbf r)\right)=0,
\end{equation}
which reduces to the Laplace equation if $\frac{h^{2}}{12 \mu}$ is a constant. In addition, the Hele-Shaw flow has the same form as the Darcy' law since $\frac{h^{2}}{12 \mu}$ behaves as the effective permeability. Similar to the work on transforming Darcy's law~\cite{PRL2011},  Eq.~(\ref{Helea}) is also mathematically form-invariant under coordinate transformations (The detailed derivation is presented in Part B of Appendix). Its form in the physical space is 
\begin{equation}\label{Helea'}
	\nabla'\cdot\left(\text{J}  \text{J}^{\top}\det \text{J}^{-1} \frac{h^{2}}{12 \mu} \nabla' p'(\mathbf r'(\mathbf r))\right)=0.
\end{equation}
If the depth keeps unchanged, we can obtain the same transformation rule for the viscosity as Eq.~(\ref{vist}). Further, we also need to verify whether the 2D model described by Eq.~(\ref{Helea'}) can be derived from anisotropic flows with a rank-4 viscosity tensor in 3D shallow cells. Based on the lubrication approximation, we can give the affirmative  conclusion. Similarly, the form of Eq.~(\ref{Hele}) in the physical space is 
$
\mathbf{v}'(\mathbf r')=-\text{J}  \text{J}^{\top}\det \text{J}^{-1}\frac{h^{2}}{12 \mu} \nabla' p',
$
and Eq.~(\ref{transv}) still holds. The detailed derivation on form invariance is offered in Part C of Appendix.  On the other hand, in Part D of Appendix, we give a rigorous analysis on how the form of the Stokes equation and the NS equations, especially the viscous stress term in them, change under curvilinear transformations. 

A more intuitive way is to use numerical simulations to check the performance of transformation devices under different flow models.
For that, we  consider a modification to Eqs.~(\ref{Helea}) and (\ref{Helea'}).
If we assume the Stokes equation is form invariant with a rank-2 viscosity tensor, the combination of Eqs.~(\ref{Stokes}) and (\ref{Hele}), i.e.,
\begin{equation}\label{anna}
	\nabla p-\nabla \cdot\left(\mu \left(\nabla \mathbf{v} +\nabla \mathbf{v} ^{\top }\right)\right)+\frac{12\mu}{h^2}\mathbf v=0,
\end{equation}
should also be form invariant
since their transformation rules are all the same. When the depth is small ($h/L\to0$), Eq.~(\ref{anna}) reduces to Eq.~(\ref{Hele}). In contrast, if $h$ is large enough, Eq.~(\ref{anna}) will reduce to Eq.~(\ref{Stokes}).
The anisotropic version of Eq.~(\ref{anna}) is
\begin{equation}\label{tanna}
	\nabla' p'-\nabla'\cdot\left({\boldsymbol {\mu}'}\cdot \left(\nabla' \mathbf{v}' +(\nabla' \mathbf{v}') ^{\top }\right)\right)+\frac{12}{h^2}{\boldsymbol {\mu}'}\cdot\mathbf v'=0.
\end{equation}
Therefore, we can deal with Hele-Shaw flows and  more general Stokes flows in a unified 2D model by considering Eqs.~(\ref{anna}) and (\ref{tanna}) with different values of $h/L$. This also helps alleviate the huge demands on computing resources for 3D simulations.  If we can construct an example that changes the form of Eq.~(\ref{anna}) under coordinate transformations when $h/L$ is not very small,  we can prove that TO is not valid in general Stokes flows.

\subsection{Numerical verification}

Numerical simulations are performed to verify our assertions about the form invariance of fluid mechanisms equations.
We use the Coefficient Form PDE module in  the commercial finite-element software  COMSOL Multiphysics (https://www.comsol.com/) to establish the anisotropic Hele-Shaw model. 
The 2D model has the same geometry and boundary conditions as Fig.~\ref{f1}(b).
In simulations, we set $L=10$~mm and apply a pressure bias $\Delta p$ on the $x$ direction. The properties of the background material  are referenced to water, including the viscosity $\mu=10^{-3}$~Pa~s and the density $\rho=1000$~kg m$^{-3}$. We use three commonly used transformation devices, i.e., the cloak [Fig.~\ref{f1}(c)], the rotator [Fig.~\ref{f1}(d)], and the concentrator [Fig.~\ref{f1}(e)].

\begin{figure}[!ht]
	\centering
	\includegraphics[width=0.8\linewidth]{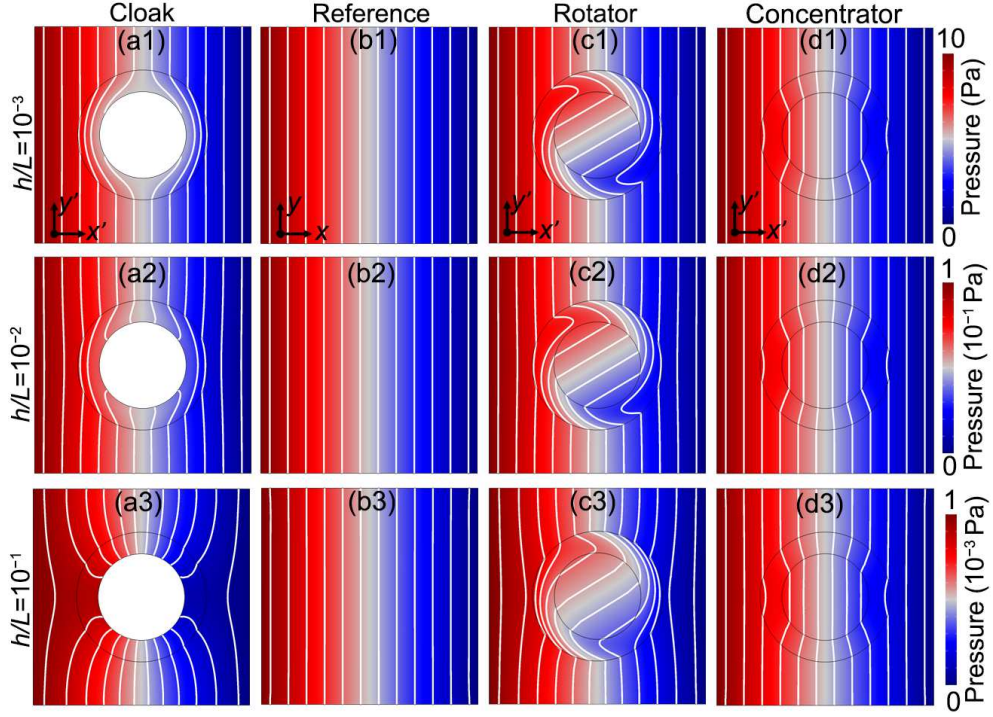}
	\caption{Simulation results for  cells with different depths. (a1)--(a3) show the pressure distributions of the cloak when $h/L$ ($\Delta p$) is $10^{-3}$ (10~Pa), $10^{-2}$ ($10^{-1}$~Pa) and $10^{-1}$ ($10^{-3}$~Pa), respectively.  The white lines are isobars. (b1)--(b3)/(c1)--(c3)/(d1)--(d3) are the corresponding pressure distributions of the reference/rotator/concentrator.}\label{f3}
\end{figure}

First, we verify the performance of hydrodynamic cloaks. 
Let $(r,\theta)$ and $(r',\theta')$ denote the radius and azimuth of the polar coordinates in the virtual and physical spaces, respectively.
 The corresponding transformation~\cite{Science06-01} is $r'=R_{1}+\frac{R_{2}-R_{1}}{R_{2}}r$~$(0<r<R_{1})$ [Figs.~\ref{f1}(b) and \ref{f1}(c)]. The cloak is the shell
$S=\{\mathbf r':R_1\leqslant r'\leqslant R_2\}$ which prevents the obstacle in it from disturbing the outside flows. Figs.~\ref{f3}(a1)--\ref{f3}(a3) present the pressure distributions when the cloak is put in cells with different depths. The region inside the cloak is a solid obstacle so it can be removed from the simulated domain with nonslip condition on $\{\mathbf r': r'=R_1\}$. For comparison, the corresponding reference pressure distributions in the virtual space are also illustrated in Figs.~\ref{f3}(b1)--\ref{f3}(b3). Here, we take $R_1=2$~mm and $R_2=3$~mm, and
let $\Delta p$ vary with  $h$ or  $h/L$ so that the Reynolds numbers (Re) can be close in all the cases. In fact, we have $\text{Re}\approx8.3\times10^{-5}$ for the reference in the following simulations if we use $h$ as the characteristic length. This small value of Re ensures that the flow is still a creeping  one even if Re is increased in cases with a cloak. 
The pressure patterns in Figs.~\ref{f3}(b1)--\ref{f3}(b3) are visually indistinguishable while those in Figs.~\ref{f3}(a1)--\ref{f3}(a3) are quite different.
The cloak with regard to $h/L=10^{-3}$ [Fig.~\ref{f3}(a1)] has a background pressure pattern looking no different from its counterparts in  Fig.~\ref{f3}(b1).
 In Fig.~\ref{f3}(a2) with a larger depth, we can see the isobars near the cloak are slightly curved. When $h/L$ increases to $10^{-1}$ in Fig.~\ref{f3}(a3), the isobars are sharply bent, and the obstacle can be easily detected. We can see the shallow geometry is necessary to maintain invisibility. When the  stress tensor term is neglected, i.e., using the real Hele-Shaw equation, we can obtain a nearly perfect cloak for any value of $h/L$ in simulations, although the corresponding 3D models may no longer be called shallow. In this cases, all the pressure distributions should have the same pattern close to Fig.~\ref{f3}(a1). Here, we say nearly perfect because the nonslip condition requires zero velocity at $\{\mathbf r': r'=R_1\}$. This conflicts with Eq.~(\ref{transv}), which does not prohibit tangential flow around the obstacle. In Fig.~\ref{f3}(a3), both the changes of equation forms and the nonslip boundary contribute to the failure of cloaking.

Further, we check the performance of another two invisible devices without introducing extra nonslip boundaries.
 Figs.~\ref{f3}(c1)--\ref{f3}(c3) present the simulated pressure distributions of the (invisible) rotator. Its transformation  [Figs.~\ref{f1}(b) and \ref{f1}(d)] is 
 	$\theta'=\theta+\theta_{0}$ $(r<R_{1})$ and
 		$\theta^{'}=a r+b$ $(R_{1}\leqslant r\leqslant R_{2})$,
 where
 $ a=\frac{\theta_{0}}{R_{1}-R_{2}} $,
 $ b=\theta+ \frac{R_{2}}{R_{2}-R_{1}}\theta_{0}$~\cite{ChenAPL07}. The rotator itself is still the shell $S$ and the physical fields inside it are expected to be rotated by $\theta_0$ (counterclockwise for a positive value), which is $-60\degree$ in  Figs.~\ref{f3}(c1)--\ref{f3}(c3).
 Also, the region outside it undergoes an identity transformation so the pressure distribution should keep invariant (invisibility). Similar to the cloak, the first two rotators in Figs.~\ref{f3}(c1) and \ref{f3}(c2) with shallower geometries work better than that in  Fig.~\ref{f3}(c3). Another typical device in TO is the (invisible) concentrator in which the physical fields can be amplified without changing the outside flows. Its
 transformation  [Figs.~\ref{f1}(b) and \ref{f1}(e)] is 
 	$r'=\frac{R_{1}}{R_{3}}r$~$(r<R_{3})$ and
 	$r'=\frac{R_{1}-R_{3}}{R_{2}-R_{3}} R_{2}+\frac{R_{2}-R_{1}}{R_{2}-R_{3}}r$~$(R_{3}\leqslant r\leqslant R_{2})$,
 where $R_3$ is an auxiliary parameter that determines the concentrating effect~\cite{RahmPNFA08}. Here we take $R_3=2.5$~mm.
 The corresponding simulation results are shown in Figs.~\ref{f3}(d1)--\ref{f3}(d3), demonstrating denser isobars in the center. It is interesting that when $h/L$ reaches $10^{-1}$, the uniform pressure distribution outside the concentrator hasn't been distorted much. In fact, it needs a quite large $h/L$ to see the obvious failure of
 invisible concentrator. Anyway, we can conclude that Eq.~(\ref{anna}) cannot be transformed to Eq.~(\ref{tanna}). At the same time, we can confirm the form invariance of the Hele-Shaw flows. Thus, the only reasonable explanation for Fig.~\ref{f3} is that the Stokes flows are not form invariant under coordinate transformations. If we add the inertial term back, the pressure distributions will be almost the same as those in Fig.~\ref{f3} due to the creeping nature. So the general laminar flows governed by the NS equations are not form invariant either.

\begin{figure}[!ht]
	\centering
	\includegraphics[width=0.8\linewidth]{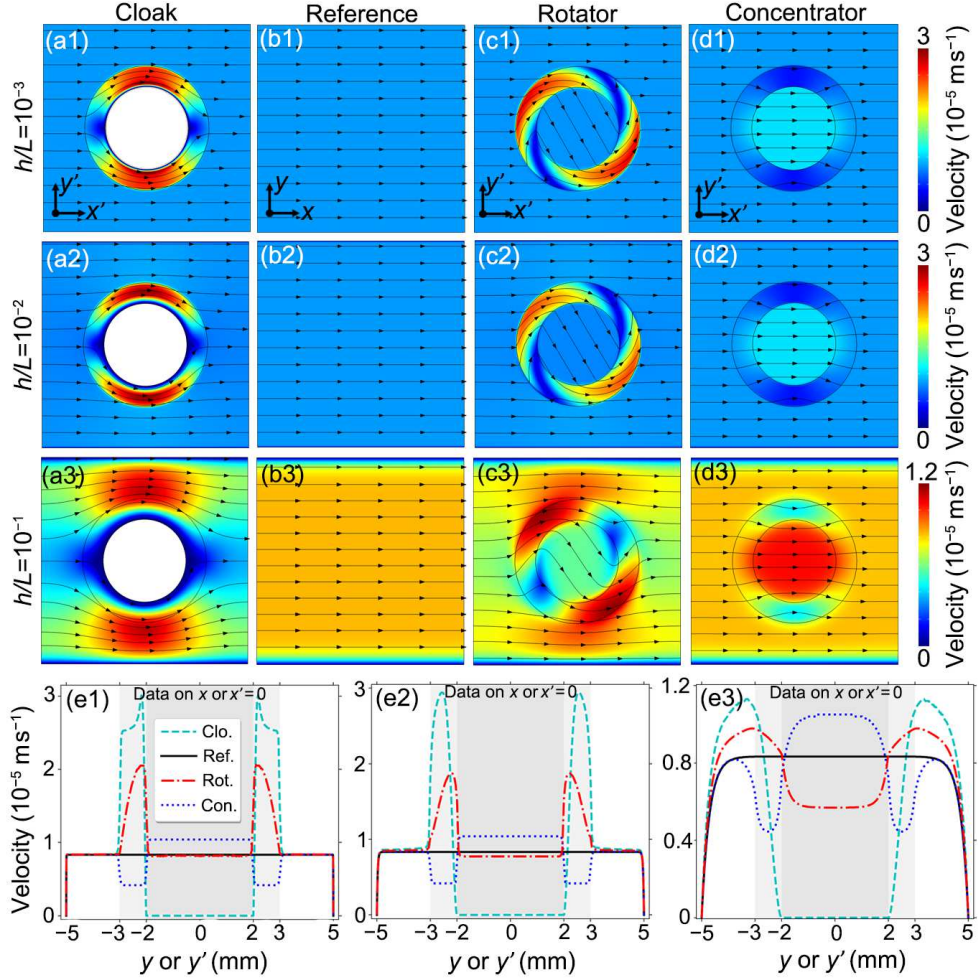}
	\caption{Simulated velocity in cells with different depths. (a1)--(a3) Pressure distributions of the cloak when $h/L$ is $10^{-3}$, $10^{-2}$ and $10^{-1}$, respectively. Different colors represent the magnitude of velocity. The black lines with arrows are streamlines. (b1)--(b3)/(c1)--(c3)/(d1)--(d3) The corresponding velocity distributions of the reference/rotator/concentrator. (e1)--(e3) Speed data read from the line $x=0$ for the devices or $x'=0$ for the reference from (a1)--(d3): (e1) $h/L=10^{-3}$; (e2) $h/L=10^{-2}$; (e3) $h/L=10^{-1}$. The different shades behind the plotted lines represent the background area (white), the device (lighter gray), and the inside of the device (darker gray).}\label{f5}
\end{figure}

Velocity distribution [Fig.~\ref{f5}] provides another perspective for judging and analyzing the validation of TO in different cells.
By comparing the background velocity (invisibility) [Figs.~\ref{f5}(a1)--\ref{f5}(d3)] and checking the uniformness inside the rotators [Figs.~\ref{f5}(c1)--\ref{f5}(c3)] and concentrators [Figs.~\ref{f5}(d1)--\ref{f5}(d3)], we can obtain the same conclusion as that from comparing pressure in Fig.~\ref{f3}.
 In particular, Fig.~\ref{f5}(d3) shows the deviation of the concentrator from the ideal effect more clearly than Fig.~\ref{f3}(d3).
Also, we can see the viscous drag plays an important role in the failure of invisible devices. Besides the inner edge of the cloak, non-slip also exists on the side walls of the cell [Fig.~\ref{f1}(b)].
 In Figs.~\ref{f5}(b1)--\ref{f5}(b3), the boundary layer becomes thicker as $h/L$ increases in the virtual space. The flows in Figs.~\ref{f5}(b1) and \ref{f5}(b2) are approximately  uniform while that in Fig.~\ref{f5}(b3) has an obvious gradient along the direction of $y$-axis near the nonslip boundaries. 
The latter case contradicts the lubrication approximation, which requires the velocity gradient in the $x$-$y$ plane to be negligible compared to that along the direction of $z$-axis.  Such boundary layer features are also present in the three devices composed of  hydrodynamic metamaterials. More specifically, we plot the corresponding speed on the line $x=0$ for the reference or $x'=0$ for the devices in Figs.~\ref{f5}(e1)--\ref{f5}(e3), which intuitively show the size of the boundary layers.

Based on our analysis in Part D of Appendix, some constraints that can make the form of Stokes flows transformation-invariant include (1) an isotropic homogeneous virtual space that is drag-free
and (2) a pressure distribution satisfying Eq.~(\ref{zj}) (differing from Eq.~(\ref{Helea'}) by a coefficient before the pressure gradient) in the physical space at the same time. It is easy to see that both of these conditions no longer hold when the lubrication approximation fails. 
In addition, the relatively better performance of the concentrator in deep cells can be explained by the less space distortion caused by its transformation. 
In Part E of Appendix, we do another set of simulations with different parameters for the three devices and  quantify their invisibility effects together with the results in Figs.~\ref{f3} and \ref{f5}. 
By adjusting $R_3$ for a larger concentrating effect, we observe a more obvious failure of the concentrator at $h/L=10^{-1}$. 
In conclusion, the rigorous establishment of TO needs  lubrication approximation  although some devices can still have a well performance in non-shallow flows if the geometric transformations only cause small space distortions.

\section{Multilayered structure with spatial variance}

The devices designed by TO need anisotropic inhomogeneous viscosities, which are quite difficult to fabricate. 
A common technique to simplify the fabrication in metamaterials is  discretizing the space and then achieving a specific isotropic (or diagonally anisotropic) and uniform material property in each unit~\cite{oe07,PRB2008,PRL2012con}. For example, 
the  cloak can be realized by multilayered structures composed of two bulk materials: one has a higher  and the other has a lower viscosity than the background.
Further, it is still not easy to flexibly increase or decrease the scalar shear viscosity. External factors such as magnetic field~\cite{pre2011} and temperature~\cite{prap2021sb} can change the viscosity in a wide range, but this will make the system more complicated. Previous studies~\cite{PRL2019,EML2020-2,PRAP2019}  put pillars into the fluid, which is an effective method to increase the viscosity. However, this technique can't reduce the viscosity so pillars were also put into the background region  to match the parameters. 
On the other hand, putting in pillars with the same height as the cell actually reduces the fluid depth to zero. So, we can use a cell with a spatially varying depth to avoid additional adjustments to the background flow. 
This also means that we can go back to the realistic 3D model. Actually, in metamaterials controlling water waves, engineering depth has been an important technique~\cite{hs,PRL2021zj} and even anisotropic depth can be realized~\cite{epl09,prl2022hy}.
Due to the variable flow cross section, the law of continuity in the physical space needs to be modified as~\cite{PF2013,pre2013,JFM2014}
\begin{equation}\label{connti2}
\nabla'\cdot(h'(x',y')\mathbf{v'})=0,
\end{equation}
and then Eq.~(\ref{Helea'}) should be generalized to
\begin{equation}\label{Heledepth}
\nabla'\cdot\left(\frac{(h')^{3}}{12 \mu} \nabla' p'\right)=0.
\end{equation}
It's easy to check  Eqs.~(\ref{connti2}) and (\ref{Heledepth}) are mathematically form-invariant, although  how the transformation matrix $\text{J}  \text{J}^{\top}\det \text{J}^{-1}$ acts on $h$ is a bit tricky. Nevertheless, we can still try to use multilayered structures to approximately mimic anisotropy.

\begin{figure}[!ht]
	\centering
	\includegraphics[width=0.9\linewidth]{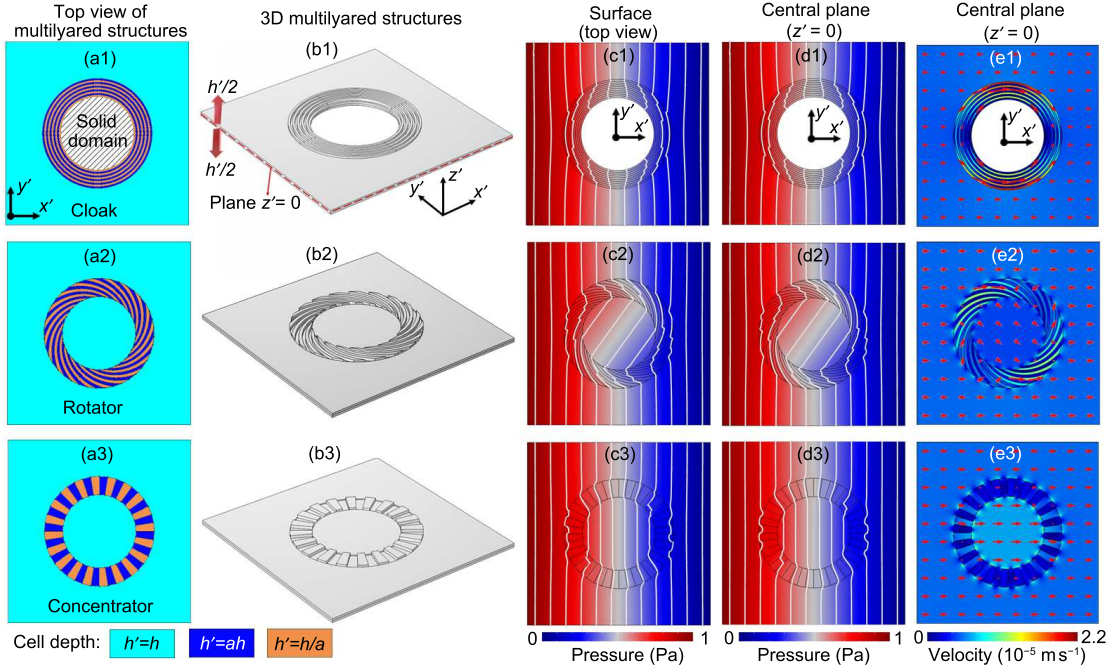}
	\caption{Structures of multilayered metamaterials with variable depth and their simulation results. (a1)--(a3) Top view (or 2D model) of the devices. The regions illustrated by different colors have different depth $h'$ in the $z'$ direction. (a1) The cloak has 10 annular layers with the same difference between inner and outer radii. (a2) The rotator has 40 layers whose helical boundaries are determined by $x'=R_1\exp(s)\cos(ks+n\pi/20)$ and $y'=R_1\exp(s)\sin(ks+n\pi/20)$, where $s\in\left[0,\,\ln\left(R_2/R_1\right)\right]$,  $k=-\theta_0/(\ln(R_2/R_1))$ and $n$ is an integer from 1 to 40~\cite{PRB2008}.
		(a3) The concentrator has 40 layers with symmetry of a $9\degree$ rotation.
		(b1)--(b3) Side views of the 3D multilayered devices. They all have a symmetry to the $z'=0$ plane, which can be generated by extending the 2D model (plane $z'=0$) by $h'/2$ in both the positive and negative directions of the $z'$-axis.   (c1)--(c3) are the pressure distributions on the surface (top view, i.e., the upper boundary $z'=h'/2$).  (d1)--(d3) are the corresponding pressure distributions on the central plane $z'=0$. (e1)--(e3) illustrate the horizontal components of the velocity vector on $z'=0$. The red arrows show the direction of horizontal velocity. Their length and the colors show the magnitude of horizontal velocity. We set $L=D=10$~mm, $h/L=2\times 10^{-3}$ and $a=2$,  and apply a pressure bias equal to 1~Pa along the $x'$-axis.}\label{f8}
\end{figure}

Fig.~\ref{f8} shows the schematic diagram of  the multilayered structures. For the cloak, the annular shell in the plane $z'=0$ 
is divided into ten concentric annuli [Fig.~\ref{f8}(a1)] and the 3D structure is generated by extruding the plane in both directions of the $z'$-axis by $h'(x',y')/2$ [Fig.~\ref{f8}(a2)].
In each layer (annulus), $h'$ is a constant. The product of $h'$ of two adjacent layers is equal to $h^2$.
 Similarly, the 3D rotator [Figs.~\ref{f8}(a2) and \ref{f8}(b2)]  or concentrator [Figs.~\ref{f8}(a3) and \ref{f8}(b3)] can also be designed by dividing the shell into helical or  rotationally symmetrical  multilayered structures. Figure.~\ref{f8} also confirms our design by 3D numerical modeling. For simplicity, we show the (flattened) pressure distributions on the upper surface of the cell in Figs.~\ref{f8}(c1)--\ref{f8}(c3)   and those on the central plane $z'=0$ in Figs.~\ref{f8}(d1)--\ref{f8}(d3). The difference of the pressure distributions on the surface and the central horizontal section of the same device is actually very small. Figures.~\ref{f8}(e1)--\ref{f8}(e3) illustrate the horizontal velocity  distributions on plane $z'=0$. 
We can see the demonstrated pressure and velocity distributions indicate the good invisibility performance
of  multilayered structures, although the isobars for the cloak in  Figs.~\ref{f8}(c1) and \ref{f8}(d1) are slightly bent--due to the extra boundary layer and approximation of the 3D model itself. From Fig.~\ref{f8}(e2), the velocity is turned about $38\degree$ clockwise inside the rotator, not fully reaching the theoretical value $60\degree$ clockwise. For the concentrator [Fig.~\ref{f8}(e3)], the velocity inside it  is amplified by a factor of 1.31 (calculated based on the spatial average), exceeding $R_m/R_1=1.25$. In fact, the expected concentrating effect should be $R_2/R_1$ for multilayered structures since $R_m$ doesn't affect the design~\cite{PRL2012con}.
The performance of our designs can be enhanced by using more layers, larger depth difference between two adjacent layers,  and, of course, a shallower cell.

\section{Conclusion}

In summary, we give a careful discussion on choosing a proper flow model to design hydrodynamic transformation metamaterials. 
Based on rigorous mathematical proof and numerical verification, we show that the form invariance required by TO can hold in Hele-Shaw flows, but not in general Stokes flows or laminar flows. When the viscous drag cannot be neglected, the flows in cells will deviate from the lubrication approximation. Then, the obstacles inside the cloaks become detectable, and the rotators and concentrators  can no longer achieve the desired regulation effects.
Furthermore, we propose a method to equivalently achieve the desired anisotropic viscosity of  metamaterials by spatially varying cell depth.
Our results not only clarify the formal basis for the application of TO in fluid mechanics, but also reveal a freedom of manipulating fluids in cells through spatially-varying configuration.

\section*{Acknowledgments}
We acknowledge financial support by the National Natural Science Foundation of China under Grants No. 12147169 and 12205101.

\section*{Appendix}

\setcounter{equation}{0}
\renewcommand{\theequation}{A\arabic{equation}}
\setcounter{figure}{0}
\renewcommand{\thefigure}{A\arabic{figure}}

\subsection{Can the rank-4 viscosity tensor be reduced to a rank-2 one?} 

Here we give a brief discussion on the existence of a rank-2 viscosity tensor. 
In general, the viscosity is a rank-4 tensor $\mathbb A$ and 
the constitutive relation of stress tensor is
\begin{equation}\label{fp}
	{\boldsymbol {\tau}}=\mathbb{A}:\nabla\mathbf{v}=A^{ijkl}\epsilon_{kl}\mathbf g_{i}\otimes \mathbf g_{j}.
\end{equation}
Since  $\nabla \mathbf{v}$ is a rank-2 tensor, we can remark it as $\boldsymbol{\epsilon}=\epsilon^{ij}\mathbf{g}_i\otimes\mathbf{g}_j$. Although a phenomenological  rank-2 tensor or a matrix of shear viscosity has been revealed in surface water waves~\cite{prl08} and the recent works on hydrodynamic metamaterials~\cite{PRL2019,EML2020-2,PRAP2019,EML2020-1}, it's still worth verifying the reconciliation between the two forms of viscosity.
For simplicity, we use the Cartesian coordinate system in this note
and Eq.~(\ref{fp}) can be expressed as
\begin{equation}\label{bengou}
	\tau_{ij}=A_{ijkl}\frac{\partial v_k}{\partial x_l}.
\end{equation}
For 3D flows, $A_{ijkl}$ has 81 components and the independent ones can be reduced under some common assumptions. For example, 
with the Onsager reciprocal relations ($A_{ijkl}=A_{klij}$), the rotational symmetry ($A_{ijkl}=A_{jikl}$)
and no-internal-friction condition under a pure uniform rotation ($A_{ijkl}=A_{ijlk}$)~\cite{landau-6}, only 21
independent components are left and the number is reduced from 16 to 6 for 2D flows. More details on the symmetry of viscosity can be found in Ref.~\cite{arxiv21}. It's obvious that 21 and 6 exceed the component numbers of a rank-2 tensor or matrix in 3D and 2D spaces, respectively. 

Since the rank-2 viscosity tensors in Refs.~\cite{PRL2019,EML2020-2,PRAP2019,EML2020-1} are only anisotropic in the horizontal plane,
we  consider the 2D anisotropic case  in the $x$–$y$ plane. With the three symmetries mentioned above, the viscous stress tensor can be expressed as~\cite{OM2003}
\begin{equation}
	\begin{bmatrix}
		\tau_{xx} \\
		\tau_{yy} \\
		\tau_{xy}
	\end{bmatrix}=\begin{bmatrix}
		A_1 & A_2 & A_3 \\
		A_2 & A_4 & A_5 \\
		A_3 & A_5 & A_6
	\end{bmatrix}\begin{bmatrix}
		\dot{e}_{xx} \\
		\dot{e}_{yy} \\
		\dot{e}_{xy}
	\end{bmatrix}.
\end{equation}
$\{A_1,A_2,...,A_6\}$ are the 6 independent components of $A_{ijkl}$.
Further, we only consider incompressible fluids, which gives an extra restriction for the components in strain rate: $\dot{e}_{xx}+\dot{e}_{yy}=0$. However, the number of independent components in $A_{ijkl}$ can only be reduced to 5, not 4 or less, so  
\begin{eqnarray}
	\begin{bmatrix}
		\tau_{xx}&\tau_{xy} \\
		\tau_{yx}&\tau_{yy}
	\end{bmatrix}\neq
	\begin{bmatrix}
		\mu_{xx}&\mu_{xy} \\
		\mu_{yx}&\mu_{yy}
	\end{bmatrix}\begin{bmatrix}
		\dot{e}_{xx}&\dot{e}_{xy} \\
		\dot{e}_{xy}&\dot{e}_{yy}
	\end{bmatrix}
\end{eqnarray}
for any matrix $\begin{bmatrix}
	\mu_{xy}
\end{bmatrix}$ as real valued functions of $(x,y)$. In conclusion, for general anisotropic fluids, we can’t reduce the rank-4 viscosity tensor to a rank-2 one.

\subsection{Form invariance of Hele-Shaw flows}

\begin{figure}[!ht]
	\centering
	\includegraphics[width=0.8\linewidth]{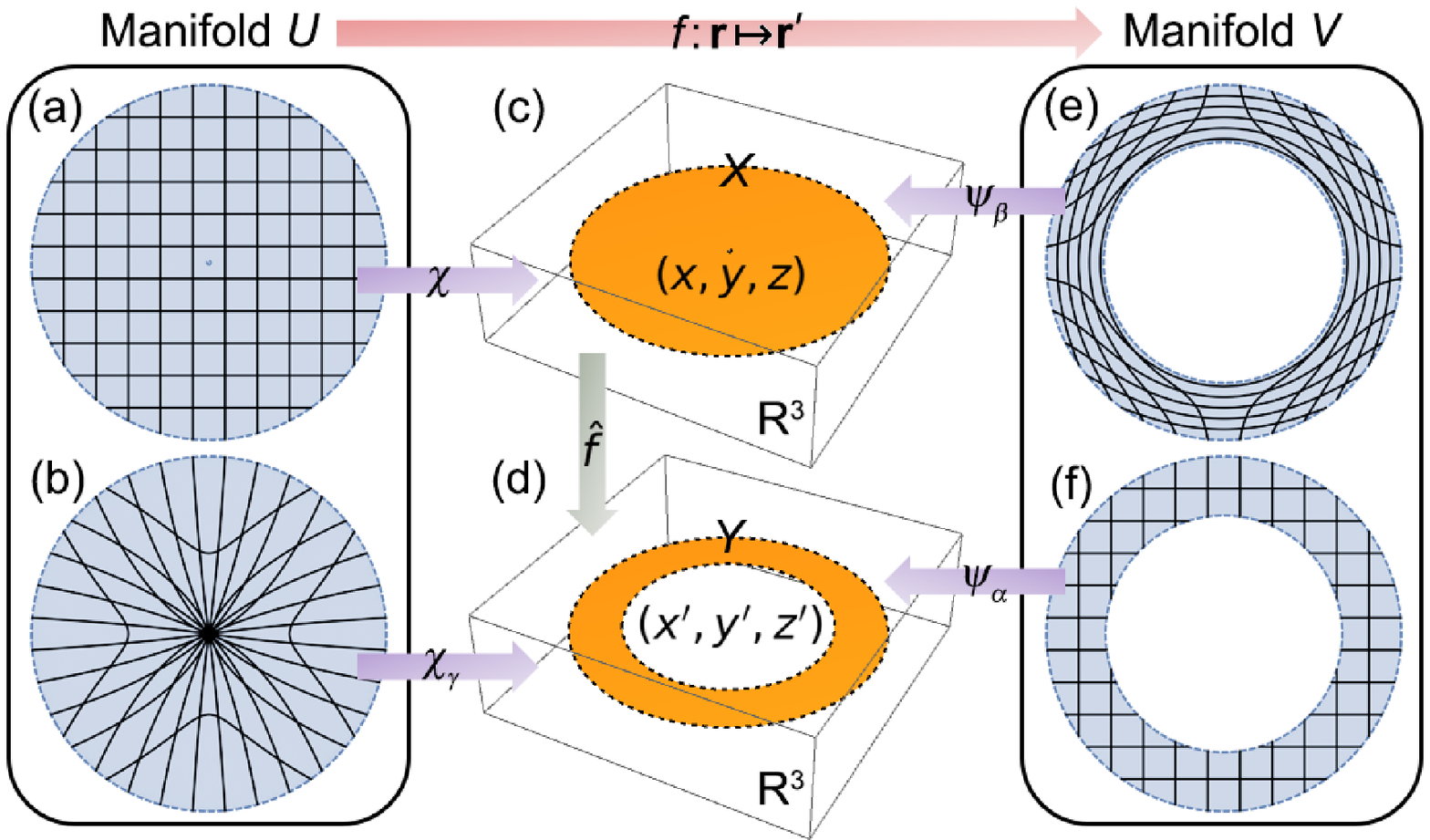}
	\caption{Illustration of the ways to understand the form invariance of TO. (a) and (b) both represent manifold $U$. The different meshes corresponding to different coordinate maps. (c) and (d) are value ranges  in $\mathsf R^3$, linked by the map $\hat{f}$. (e) and (f) both represent manifold $V$. $f$ is the map (diffeomorphism) from $U$ to $V$. We actually take the cloak as an example.
		(c) gives the value range of chart $(U,\chi)$ and $(V,\psi_{\beta})$. (d) gives the value range of chart $(U,\chi_{\gamma})$ and $(V,\psi_{\alpha})$. Form invariance means the governing equations written by the coordinates in (a) and (e), or (b) and (f), are the same.}\label{s1}
\end{figure}

In this note, we give a detailed proof for the form invariance of Hele-Shaw flows.
As the starting point, we consider a  map  $f$ in 3D Euclidean space $\mathsf{E}^3$: $U \to V$ ($U,V\subseteq\mathsf{E}^3$) or $\mathbf r \mapsto \mathbf r'$ ($\mathbf r\in U,\mathbf r'\in V$) [Fig.~\ref{s1}]. $U$ (the virtual space) and $V$ (the physical space) can seen as two sub-manifolds of the simple manifold $\mathsf{E}^3$.
The position vectors can be expressed by the global Cartesian coordinates in $\mathsf{E}^3$, i.e., $\mathbf {r}=x\mathbf e_{x}+y\mathbf e_{y}+z\mathbf e_{z}$ and 
$\mathbf r'=f(\mathbf {r})=x'\mathbf e_{x}+y'\mathbf e_{y}+z'\mathbf e_{z}.$ Here $\{\mathbf e_{x},\mathbf e_{y},\mathbf e_{z}\}$ is  the orthonormal basis. Then we can see $f$ has the same effect as another map $\hat{f}$ between two subsets $X,Y$ in real 3-space $\mathsf {R} ^{3}$:
\begin{equation}
	{\displaystyle {\begin{cases}\hat{f}:X\subseteq\mathsf {R} ^{3}\to Y\subseteq\mathsf {R} ^{3},\\(x,y,z)\mapsto (x',y',z').\end{cases}}}
\end{equation}
Here $X=\chi(U)$, $Y= \psi_{\alpha}(V)$ and $\hat{f}=\psi_{\alpha}\circ( f\circ\chi^{-1})$. $\chi$ and $ \psi_{\alpha}$ are both bijections (or more exactly, homeomorphisms) that use only one coordinate map  to give the coordinate of every point in each manifold, i.e., the atlas~\cite{jost}. The pairs of manifold and coordinate map, i.e., (coordinate) charts, can be denoted as $(U,\chi)$ [Figs.~\ref{s1}(a) and \ref{s1}(c)]  and $(V, \psi_{\alpha})$ [Figs.~\ref{s1}(f) and \ref{s1}(d)] .
For TO, $f$ must be a bijection and thus its inverse $f^{-1}$ exists. Further, both $f$ and $f^{-1}$ should be smooth enough so  $f$ and $\hat{f}$ are diffeomorphisms. In this way, Jacobian $\text{J}_{x'x}=\partial x/\partial x'$ and its inverse are both well-defined.

First, the law of continuity [Eq.(1)] in $U$ under a curvilinear coordinate system of $\mathsf{E}^3$ can be written as
\begin{equation}\label{niubia}
	\nabla\cdot(\rho \mathbf v)=\frac{1}{\sqrt{g}}\partial_i\left(\sqrt{g} \rho v^i\right)=0.
\end{equation}
Here $v^i$ is the contravariant component of velocity using contravariant coordinates $\{x^i,x^j,x^k\}$ and the corresponding basis $\{\mathbf g_i,\mathbf g_j,\mathbf g_k\}$.
$g$ is the determinant of the metric matrix $g_{ij}=\mathbf g_i\cdot \mathbf g_j$ which satisfies $\sqrt{g}=\frac{1}{\operatorname{det} \operatorname{J}}$. The Jacobian matrix is  $\operatorname{J}_{ix}=\partial x_i/\partial x$, with regard to the coordinate transformation  from $\{x,y,z\}$ to $\{x^i,x^j,x^k\}$. The form of Eq.~(\ref{niubia}) in arbitrary  curvilinear coordinate system  differs by only the value of $g$. This is the basic requirement for building a TO theory.
In particular, we can choose a coordinate system which makes $\partial x_i/\partial x=\partial x'/\partial x$ since $f$ is a diffeomorphism. In fact, using coordinates $\{x^i,x^j,x^k\}$ means constructing another atlas with a single chart for $U$ (denoted by $(U,\chi_{\gamma})$), and there must exist a homeomorphism satisfying $\chi_{\gamma}=\hat{f}\circ\chi=\psi_{\alpha}\circ f$. Also, it's easy to see $\chi_{\gamma}(U)=Y$ [Figs.~\ref{s1}(b) and \ref{s1}(d)]. Then 
we can obtain
\begin{equation}\label{niubia2}
	\partial_{x'}\left( \dfrac{1}{\operatorname{det} \operatorname{J}}\rho v^{x'}(x',y',z')\right)=0
\end{equation}
from Eq.~(\ref{niubia}). With $v^{x'} (x',y',z')\text{J}^{-1}_{xx'} =v^{x}(x,y,z)$, Eq.~(\ref{niubia2}) can be rearranged as the law of continuity in $V$ [Eq.~(5)]
and the transformation rule for velocity [Eq.~(6)].
The key point here is that Eq.~(\ref{niubia}) can describe the law of continuity in different manifolds using the charts  $(U,\chi_{\gamma})$ and $(V,\psi_{\alpha})$ at the same time, which are linked by the  diffeomorphism $f$.  Actually, the form invariance required by TO  in previous studies~\cite{Science06-01} was usually explained by another two charts including $(U,\chi)$. The left chart we haven't mentioned is $(V,\psi_{\beta})$, where $\psi_{\beta}=\hat{f}^{-1}\circ \psi_{\alpha}$ [Figs.~\ref{s1}(e) and \ref{s1}(c)]. We have $\psi_{\beta}(V)=X=\chi(U)$ and $\chi=\psi_{\beta}\circ f$. If the law of continuity written by $(V,\psi_{\beta})$ is exactly the same as its counterpart by $(U,\chi)$, we can say it is form-invariant. By doing a coordinate transformation from $(V,\psi_{\beta})$ to the partial Cartesian coordinates  $(V,\psi_{\alpha})$, the transformation rules for velocity can be obtained again. In fact, Figs.~\ref{f1}(b)--\ref{f1}(e) are drawn based on  the latter way of understanding.

Second, we  observe  the  form of Hele-Shaw equation [Eq.~(\ref{Hele})] in $U$ using contravariant coordinates $\{x^i,x^j,x^k\}$:
\begin{equation}\label{Darcy}
	\left(\dfrac{h^2}{12}(\partial_i p)g^{ij}+\mu v^j \right)\mathbf g_j=0.
\end{equation}
Do the same procedure as we have done to the law of continuity and we can obtain
\begin{equation}\label{Darcy2}
	\dfrac{1}{	\operatorname{det}\text{J}}	\left(\dfrac{h^2}{12}(\partial_{y'} p'(x',y',z'))\text{J}_{y'x}\text{J}_{xx'}^{\top}+\mu v^{x'}  \right)\mathbf e_{x}=0
\end{equation}
for $(U,\chi_{\gamma})$ and $(V,\psi_{\alpha})$, which indicates the form invariance. The pressure in   $V$ should satisfy
\begin{equation}\label{mubiao2}
	p'(\mathbf r')=p(f^{-1}(\mathbf r')),
\end{equation}
and the viscosity in $V$ should be transformed to
$
\mu'_{x'y'}(\mathbf r')=\mu\det\text{J}\text{J}^{-\top}_{x'x}\text{J}^{-1}_{xy'}
$ [Eq.~(7)].
Another approach is directly using Eq.~(10):
\begin{equation}\label{trans-heat-2}
	\nabla \cdot \left(\dfrac{h^2}{12\mu}\nabla p\right)=\frac{1}{\sqrt{g}}\partial_j\left(\sqrt{g} \dfrac{h^2}{12\mu} g^{ij}\partial_i p\right).
\end{equation}
With $g^{ij}=\text{J}_{ix}\text{J}_{xj}^{\top}$, we can obtain  Eq.~(7) again.

If we directly consider the form invariance of the 3D model described by $
\mathbf{v}=-\frac{h^{2}}{8 \mu}\left(1-\left(\frac{2 z}{h}\right)^{2}\right) \nabla p
$ or $\nabla\cdot\left(\frac{h^{2}}{8 \mu}\left(1-\left(\frac{2 z}{h}\right)^{2}\right) \nabla p\right)=0$, we can get the same conclusion as 
Eq.~(10) since they  differ only by a factor along the $z$-axis. Of course, this factor  contains $z$. If the coordinate transformation only happens in the horizontal plane, the 3D and 2D models can share
the same streamlines in the  horizontal plane and the devices we design, i.e., the cloak, the rotator, and the concentrator, can still work in the 3D model.

\subsection{Anisotropic and inhomogeneous Hele-Shaw flows.}

We have known the Hele-Shaw equation is form-invariant under coordinate transformations and the transformation rule requires a rank-2 viscosity tensor. In this note , we show that Eq.~(\ref{Helea'}) is indeed the governing equation for anisotropic inhomogeneous Stokes flows in shallow cells.
Under lubrication approximation, the strain rate with the Cartesian coordinates is
\begin{equation}
	\left(\nabla \mathbf{v}+\nabla \mathbf{v}^{\top}\right) \simeq \begin{bmatrix}{0} & {0} & {\frac{\partial v_x}{\partial z}} \\ {0} & {0} & {\frac{\partial v_y}{\partial z}} \\ {\frac{\partial v_x}{\partial z}} & {\frac{\partial v_y}{\partial z}} & {0}\end{bmatrix},
\end{equation}
which comes from the fact that
$
\frac{\partial}{\partial x} \ll \frac{\partial}{\partial z}$ and $\frac{\partial}{\partial y} \ll \frac{\partial}{\partial z}
$~\cite{InFlow} 
since the scale in the direction of $z$-axis is much smaller than those in the $x$--$y$ plane. Similarly, the Stokes equation [Eq.~(\ref{Stokes})] and the law of continuity [Eq.~(\ref{contunuity})] can respectively be reduced to~\cite{InFlow}  
\begin{equation}\label{shirun}
	\frac{\partial p}{\partial x}=\frac{\partial \tau_{xz}}{\partial z}, \quad \frac{\partial p}{\partial y}=\frac{\partial \tau_{yz}}{\partial z}, \quad \frac{\partial p}{\partial z}=\frac{\partial \tau_{zz}}{\partial z}=0, 
\end{equation}
and
\begin{equation}\label{shirun2}
	\frac{\partial v_x}{\partial x}+\frac{\partial v_y}{\partial y}=0.
\end{equation}
Now we consider the case when the shear viscosity is anisotropic and inhomogeneous in the $x$--$y$ plane, which can be expressed as
\begin{eqnarray}\label{shirun3}
	\begin{bmatrix}
		\mu_{xy}
	\end{bmatrix}=\begin{bmatrix}
		\mu_{xx}(x,y)&\mu_{xy}(x,y)&0 \\
		\mu_{yx}(x,y)&\mu_{yy}(x,y)&0\\
		0&0&\mu_{zz}
	\end{bmatrix}.
\end{eqnarray} 
Then the stress is
\begin{eqnarray}\label{shirun4}
	\begin{bmatrix}
		\tau_{xy}
	\end{bmatrix}=\begin{bmatrix}
		0&0&\tau_{xz} \\
		0&0&\tau_{yz}\\
		\tau_{zx}&\tau_{zy}&0
	\end{bmatrix}=\begin{bmatrix}
		0&0&\mu_{xx}\frac{\partial v_x}{\partial z}+\mu_{xy}\frac{\partial v_y}{\partial z} \\
		0&0&\mu_{yx}\frac{\partial v_x}{\partial z}+\mu_{yy}\frac{\partial v_y}{\partial z} \\
		\mu_{zz}\frac{\partial v_x}{\partial z}&\mu_{zz}\frac{\partial v_y}{\partial z}&0
	\end{bmatrix}.
\end{eqnarray} 
We can notice that the pressure is only the function of $x$ and $y$ (recall $\tau_{zz}$=0). Do an integral in Eq.~(\ref{shirun}) and we have
\begin{eqnarray}
	\begin{bmatrix}
		\mu_{xx}(x,y)&\mu_{xy}(x,y) \\
		\mu_{yx}(x,y)&\mu_{yy}(x,y)
	\end{bmatrix}
	\begin{bmatrix}
		\frac{\mathrm{d} v_x}{\mathrm{d} z}  \\
		\frac{\mathrm{d} v_y}{\mathrm{d} z}
	\end{bmatrix}  =  \begin{bmatrix}
		\frac{\partial p}{\partial x}  \\
		\frac{\partial p}{\partial y}
	\end{bmatrix}z.
\end{eqnarray}
Finally we obtain the velocity distribution in the $x$--$y$ plane:
\begin{eqnarray}
	\begin{bmatrix}
		v_x  \\
		v_y
	\end{bmatrix}  = -\frac{h^{2}}{8 }\left(1-\left(\frac{2 z}{h}\right)^{2}\right)\begin{bmatrix}
		\mu_{xx}(x,y)&\mu_{xy}(x,y) \\
		\mu_{yx}(x,y)&\mu_{yy}(x,y)
	\end{bmatrix}^{-1} \begin{bmatrix}
		\frac{\partial p}{\partial x}  \\
		\frac{\partial p}{\partial y}
	\end{bmatrix},
\end{eqnarray}
and the average values along the $z$ axis:
\begin{eqnarray}
	\begin{bmatrix}
		\overline v_x  \\
		\overline v_y
	\end{bmatrix}  =-\frac{h^{2}}{12 }\begin{bmatrix}
		\mu_{xx}(x,y)&\mu_{xy}(x,y) \\
		\mu_{yx}(x,y)&\mu_{yy}(x,y)
	\end{bmatrix}^{-1} \begin{bmatrix}
		\frac{\partial p}{\partial x}  \\
		\frac{\partial p}{\partial y}
	\end{bmatrix}.
\end{eqnarray}
In particular, when the shear viscosity is a scalar, we obtain the familiar results standing for isotropic Hele-Shaw flows [Eq.~(\ref{Hele})].

Here we use the anisotropic rank-2 shear viscosity in the derivations, which can be proven consistent with the anisotropic rank-4 viscosity tensor $A_{ijkl}$. Under the lubrication approximation, a simple form of $\tau_{xz}$ and $\tau_{yz}$ and can be obtained, writing
\begin{equation}
	\tau_{xz}=A_{xzxz}\frac{\partial v_x}{\partial z}+A_{xzyz}\frac{\partial v_y}{\partial z}, \quad \tau_{yz}=A_{yzxz}\frac{\partial v_x}{\partial z}+A_{yzyz}\frac{\partial v_y}{\partial z}.
\end{equation}
If we take
\begin{eqnarray}
	\begin{bmatrix}
		\mu_{xx}&\mu_{xy} \\
		\mu_{yx}&\mu_{yy}
	\end{bmatrix}=
	\begin{bmatrix}
		A_{xzxz}&A_{xzyz} \\
		A_{yzxz}&A_{yzyz}
	\end{bmatrix},
\end{eqnarray}
the rank-4 viscosity can be reduced to the rank-2 shear viscosity. The bulk viscosity disappears here due to the incompressibility  [Eq.~(\ref{shirun2})]. The zero value of $\tau_{zz}=A_{zzxz}\frac{\partial v_x}{\partial z}+A_{zzyz}\frac{\partial v_y}{\partial z}$ can be guaranteed by taking $A_{zzxz}=A_{zzyz}=0$. A similar discussion on constructing such a ``generalized Darcy's law'' can also be found in a recent work~\cite{arxiv22}, which further considers the effect of gravity.

\subsection{Form variance of the Stokes equation and NS equations.} 
Here we investigate how the NS equations [Eq.~(\ref{NS})] and the Stokes equation [Eq.~(\ref{Stokes})] change their form under curvilinear transformations.
There are three parts in Eq.~(\ref{NS}): the pressure term, the inertial (or advection) term, and the viscous stress term.
With contravariant basis $\left\{\mathbf{g}^i,\,\mathbf{g}^j,\,\mathbf{g}^k\right\}$ and  contravariant coordinates $\left\{x^i,\,x^j,\,x^k\right\}$, it's easy to see the pressure term is form-invariant:
\begin{equation}\label{pressure}
	\nabla p= \dfrac{\partial p}{\partial x^l}\mathbf g^{l}.
\end{equation}
The inertial term $(\mathbf {v} \cdot \nabla )\mathbf {v}$, which can be neglected in Eq.~(\ref{Stokes}), can be written as
\begin{equation}
	v^i\partial_i\left(v^j\mathbf g_j \right)=v^i\left(\partial_iv^j \right)\mathbf g_j+v^i v^j\partial_i \mathbf g_j.
\end{equation} 
The connection term $\partial_i \mathbf g_j$ is actually not the same in different curvilinear coordinate systems. If this form change can't be offset by another in the divergence of viscous stress $\nabla \cdot{{ \boldsymbol {\tau}}}$, the NS equations can't be form-invariant. 

Now we check the form of $\nabla \cdot{{ \boldsymbol {\tau}}}=\nabla \cdot \mu \left(\nabla \mathbf{v} +\nabla \mathbf{v} ^{\top }\right)$. We can deal with this question from different perspectives. 

\textbf{Approach 1.} The term $\nabla \cdot  {\boldsymbol {\tau}}$ as a whole in curvilinear coordinate systems is
\begin{equation}\label{lalala}
	\begin{aligned}
		\nabla \cdot  {\boldsymbol {\tau}} &=\mathbf g^{n} \cdot \dfrac{\partial}{\partial x^n} \left(\tau^{ij}\mathbf g_{i}\otimes\mathbf g_{j}  \right)\\
		&=\dfrac{1}{\sqrt{g}}\partial_i\left(\sqrt{g} \tau^{ij} \right)\mathbf g_{j}+\tau^{ij} \Gamma^{n}_{ij} \mathbf {g}_{n}.
	\end{aligned}
\end{equation}
Here the Christoffel symbol $\Gamma^{n}_{ij} =\left(\partial_i \mathbf{g}_j\right)\cdot\mathbf{g}^n$  varies with the choice of coordinate system. 
Obviously, it and the connection in inertial term do not cancel out.
This is the easiest way to find the general NS equations and the Stokes equation  don't meet the requirement of TO.

\textbf{Approach 2.} We assume that there is a rank-2 viscosity tensor $\boldsymbol {\mu}$, although we're not sure if it really exists.
Then we focus on the generalized form of Eq.~(\ref{wbsl}): $\nabla\cdot{\boldsymbol {\tau}}=\nabla\cdot\left({\boldsymbol {\mu}}\cdot \left(\nabla \mathbf{v} +\nabla \mathbf{v} ^{\top }\right)\right)$ [Eq.~(\ref{yl})]. Recall $\nabla \mathbf{v}\triangleq\boldsymbol{\epsilon}=\epsilon^{ij}\mathbf{g}_i\otimes\mathbf{g}_j$, which is a little different from the strain rate tensor. We should notice that, in the curvilinear coordinate system, $\epsilon^{ij}$ is not simply  equal to $\frac{\partial v^l}{\partial x^j}$, but involves  Christoffel symbols $\Gamma^{i}_{kl}$:
\begin{equation}
	\boldsymbol{\epsilon}=\nabla \mathbf {v} =g^{kj}\left({\frac {\partial v^{i}}{\partial x^{k}}}+\Gamma^{i}_{kl}v^{l}\right)\mathbf {g} _{i}\otimes \mathbf {g} _{j}.
\end{equation}
Finally we can find
\begin{equation}\label{viscous}
	\nabla\cdot\left({\boldsymbol {\mu}}\cdot \left(\nabla \mathbf{v} +\nabla \mathbf{v} ^{\top }\right)\right)=\dfrac{1}{\sqrt{g}}\partial_i\left(\sqrt{g} \mu^{ij}\left(\epsilon_{jm}+\epsilon_{mj}\right) \right)\mathbf g^{m}+\left(\mu^{ij}\left(\epsilon_{jm}+\epsilon_{mj}\right)   \right)\partial_i \mathbf {g}^m.
\end{equation}
Not surprisingly, the troublesome connection term still appears, and the divergence of  viscous stress is not form-invariant.

\textbf{Approach 3.} 
We start directly from Eq.~(\ref{fp}), which is expressed by the rank-4 viscosity tensor. In fact, writing the symmetric form $\nabla \mathbf{v} +\nabla \mathbf{v} ^{\top }$ in Eq.~(\ref{viscous})  indicates $A_{ijkl}=A_{ijlk}$~\cite{landau-6}.
Again, we can rewrite the viscous stress term using tensor components:
\begin{equation}\label{lala}
	\begin{aligned}
		\nabla \cdot  {\boldsymbol {\tau}} &=\mathbf g^{n} \cdot \dfrac{\partial}{\partial x^n} \left(A^{ijkl}\epsilon_{kl}\mathbf g_{i}\otimes \mathbf g_{j} \right)\\
		&=\dfrac{1}{\sqrt{g}}\partial_i\left(\sqrt{g} A^{ijkl}\epsilon_{kl} \right)\mathbf g_{j}+A^{ijkl}\epsilon_{kl} \Gamma^{n}_{ij} \mathbf {g}_{n},
	\end{aligned}
\end{equation}
and obtain the same conclusion as the previous two approaches. 

Further, these approaches can help us find the conditions when TO can work in Stokes flows.
Since $\Gamma^{n}_{ij}=\Gamma^{n}_{ji}$, we can find one possible condition to make Eqs.~(\ref{lalala}) and (\ref{lala}) look form-invariant:
\begin{equation}\label{ijji}
	\tau^{ij}=-\tau^{ji}.
\end{equation}
For a non-trivial case where $\boldsymbol \epsilon\neq\boldsymbol 0$, we can see  Eq.~(\ref{ijji}) actually requires
\begin{equation}
	A^{ijkl}=-A^{jikl}.
\end{equation}
Under this condition, we can rewrite Eq.~(\ref{lalala}) without the connection term as
\begin{equation}\label{slw}
	\dfrac{1}{\sqrt{g}}\partial_i\left(\sqrt{g} \tau^{ij} \right)\mathbf g_{j}-\dfrac{\partial p}{\partial x^k}g^{kl}\mathbf g_{l}=0.
\end{equation}
It seems difficult to judge the form invariance directly based on Eq.~(\ref{slw}). Here, we apply the divergence operator on both sides of Eq.~(\ref{slw}) and  have
\begin{equation}\label{xiaohu}
	\partial_j\left(\sqrt{g} g^{ij}\partial_i p \right)+\partial_j\partial_i\left(\sqrt{g}\tau^{ij}\right)=0.
\end{equation}
If the flow before transformation is incompressible, isotropic ad homogeneous with a constant shear viscosity $\mu$, meaning $\tau^{ij}=\mu\epsilon^{ij}$, then Eq.~(\ref{xiaohu}) can be rearranged as
\begin{equation}\label{wei}
	\dfrac{1}{\sqrt{g}}\left(\partial_j\left( \dfrac{1}{\mu}\sqrt{g} g^{ij}\partial_i p \right)+\partial_j\partial_i\left(\sqrt{g}\epsilon^{ij}\right)\right)=0.
\end{equation}
However, if $\mu\neq0$, to satisfy Eq.~(\ref{ijji}), we must require
\begin{equation}\label{ijji2}
	\epsilon^{ij}=-\epsilon^{ji},
\end{equation}
and it's easy to find the viscous term $\partial_j\partial_i\left(\sqrt{g}\tau^{ij}\right)$ (or $\partial_j\partial_i\left(\sqrt{g}\epsilon^{ij}\right)$) in Eq.~(\ref{xiaohu}) (or Eq.~(\ref{wei})) should vanish (the so-called ``drag-free''~\cite{PRL2019}). In addition, Eq.~(\ref{ijji2}) can lead to $\epsilon_{ij}=-\epsilon_{ji}$  so the whole Eq.~(\ref{viscous}) also becomes a zero vector in this case. This is just a scenario when the pressure satisfies the Laplace's equation
\begin{equation}\label{cjb}
	\nabla\cdot\nabla p=0,
\end{equation}
or
\begin{equation}\label{cjb2}
	\nabla\cdot\left(\dfrac{1}{\mu}\nabla p\right)=0,
\end{equation}
since $\mu$ is a constant. Eq.~(\ref{cjb2}) is mathematically form-invariant, and it should be transformed to
\begin{equation}\label{zj}
	\nabla'\cdot\left(\text{J}  \text{J}^{\top}\det \text{J}^{-1}\dfrac{1}{\mu}\nabla' p'\right)=0
\end{equation}
in the physical space.
Since $g^{ij}=\text{J}_{ix}\text{J}_{xj}^{\top}$, we can obtain the transformation rule for the viscosity again.
However, to make the governing equations reduce to Eq.~(\ref{zj}), we still have to seek for the help of shallow geometries or the Darcy's law in porous media~\cite{PRL2011}.

\subsection{Deviations for invisible devices with different parameters}

In this note, we compare the performance of the cloak, the rotator and the concentrator when their corresponding transformations can have  different geometric parameters.
Since the three devices are all expected to be visible, we can use the following quantity
to measure their invisibility effect:
\begin{equation}
	\frac{\left \langle \left | p'-p \right | \right \rangle_{r'>R_2}}{\Delta p}  =\frac{\int_{r'>R_2} \left | p'-p \right |\rm{d}r'\rm{d}\theta' }{\Delta p\int_{r'>R_2}^{} \rm{d}r'\rm{d}\theta' }\approx \frac{\sum_{i=1}^{N} \left| p'-p \right | }{N\Delta p}.	
\end{equation}
The two formulas on the left side of the approximation symbol represent the average absolute difference (deviation) between the background pressures in the physical and virtual spaces. The formula on the right side gives how to calculate this deviation based on the data of numerical simulations. $N$ is the number of the nodes of finite-element method in the background area, i.e., $\{\mathbf r:r>R_2\}$ or $\{\mathbf r':r'>R_2\}$. 

\begin{table}\label{t1}
	\caption{\label{tab:table3}Average pressure deviation. We give the simulation results with the parameters in Fig.~\ref{f3} and another set of parameters.}
	\begin{ruledtabular}
		\begin{tabular}{ccccccc}
			&\multicolumn{2}{c}{Cloak}&\multicolumn{2}{c}{Rotator}&\multicolumn{2}{c}{Concentrator}\\
			$h/L$&$R_1=2$~mm&$R_1=1$~mm&$\theta=60\degree$
			&$\theta=30\degree$& $R_m=2.5$~mm&$R_m=2.9$~mm\\ \hline
			$10^{-3}$&$0.059\%$&$0.0071\%$ &$0.041\%$&$0.0056\%$&$0.0015\%$&$0.0061\%$\\
			$10^{-2}$&$0.59\%$
			&$0.072\%$&$0.31\%$&$0.060\%$&$0.0094\%$ &$0.035\%$\\
			$10^{-1}$&$4.0\%$&$0.95\%$
			&$1.7\%$& $0.51\%$&$0.10\%$&$0.65\%$\\$10^{-1}$\footnote{The governing equation Eq.~(\ref{tanna}) in $\{\boldsymbol r':R_1\leqslant r'\leqslant R_2\}$ is replaced by $\nabla' p'+\frac{12}{h^2}{\boldsymbol {\mu}'}\cdot\mathbf v'=0$ [Eq.~(\ref{anna})].}&$0.019\%$&$0.014\%$ 
			&$0.058\%$& $0.020\%$ &$0.0062\%$ &$0.034\%$\\
		\end{tabular}
	\end{ruledtabular}
\end{table}

TABLE 1 shows the calculated deviations for the three invisible devices. Each device corresponds to two columns of data, where the first column corresponds to the results in Fig.~\ref{f3}, and the second column comes from a new set of transformation parameters. For the cloak, we use a smaller $R_1$ in the new simulations. For the rotator, we reduce the rotation angle $\theta$. The two changes decrease the intensity of spatial transformations. Conversely, we use a larger $R_m$ for the new concentrator which makes the  transformation more drastic. From TABLE 1, we can see the deviations increase  with $h/L$ and some even become non-negligible, indicating the failure of TO. The new parameters reduce the deviations of the cloak and rotator while that of the new concentrator becomes obvious now at $h/L=10^{-1}$.
The last row of data gives the deviations at $h/L=10^{-1}$ when the governing equation is the really form-invariant one [Eq.~(\ref{anna})] in the device domain. 
Since $h/L=0.1$ is not very large, such a modification is enough to generate negligible deviations. When $h/L$ further becomes larger, 
we must use Eq.~(\ref{anna}) in the whole simulation domain to see the devices work well.

\end{document}